\documentclass[conference]{IEEEtran}
\usepackage{graphicx}
\usepackage{amsmath}
\usepackage{amssymb}
\usepackage{fancyhdr}
\newlength{\zeroheight}
\usepackage{float}
\usepackage{epstopdf}

\begin{document}
     
\title{\LARGE \bf Monitoring voltage collapse margin by measuring the area voltage across 
several transmission lines with synchrophasors}
\author{
\IEEEauthorblockN{ Lina Ramirez\hspace{2cm}Ian Dobson}
\IEEEauthorblockA{Electrical and Computer Engineering Dept.\\
Iowa State University\\
Ames IA USA \\
linar@iastate.edu, dobson@iastate.edu}}
\fancyhead[c]{\textnormal{\small IEEE Power and Energy Society General Meeting, July 2014, National Harbor MD USA}}
\renewcommand{\headrulewidth}{ 0.0pt}
 \fancyfoot[L]{~\\[-20pt]Preprint \copyright 2014 IEEE. \small Personal use of this material is permitted. Permission from IEEE must be obtained for all other uses, in any current or future media, including reprinting/republishing this material for advertising or promotional purposes, creating new collective works, for resale or redistribution to servers or lists, or reuse of any copyrighted component of this work in other works.}
  \fancyfoot[C]{~ }

\maketitle
\thispagestyle{fancy}
\begin{abstract}  
We consider the fast monitoring of voltage collapse margin using synchrophasor measurements at both ends of transmission lines that transfer power from two generators to two loads.
This shows a way to extend the monitoring of a radial transmission line to multiple transmission lines.
The synchrophasor voltages are combined into a single complex voltage difference across an area containing the transmission lines that can be monitored in the same way as a single transmission line.
We identify ideal conditions under which this reduction to the single line case perfectly preserves the margin to voltage collapse, and give an example that shows that the error under practical non-ideal conditions is reasonably small.\\
\end{abstract}


\section*{Notation}
\vspace{2pt}
 \begin{tabular}{ @{}ll @{}}
 $g$&Generation bus\\
 $\ell$&Load bus\\
 $TH$&Th\'evenin\\
 $V_i$&Complex voltage at bus  $i$\\
 $V_{ij}$&Complex voltage across line between buses $i$ and $j$\\
 $V_{g\ell}$&Complex voltage across the area\\
 $I_{ij}$&Complex current through line between buses $i$ and $j$\\
 $Z_{ij}$&Impedance of line between buses $i$ and $j$\\
 $Y_{ij}$&Admittance of line between buses $i$ and $j$\\
 $w_{ij}$&Weight of the line between buses $i$ and $j$\\
 $S_{i}$&Complex power at bus $i$\\
 $S_{ij}$&Complex power through line between buses $i$ and $j$\\
\end{tabular}

\section{Introduction}
\label{intro}

Inexpensive, clean and abundant generation is generally distant from the load, creating the need that some areas export large quantities of power through transmission corridors. 
Large transfer of power through the corridors increases the risk of voltage collapse and blackout, and it is useful to monitor the margin to voltage collapse, so that the operator can take prompt action to restore the margin if it becomes too small.

Of course, voltage collapse can be detected and avoided by calculations based on the state estimator \cite{CutsemVournasbook, GreeneTPS97}, but there is scope for additional monitoring based on synchrophasor measurements, particularly since the state estimation and the calculations take some time to complete (several minutes), and in some fault situations the state estimator is not reliable, and it is good to have another, independent approach available.
 \renewcommand{\footskip}{20pt}
 \pagestyle{plain}
Since the 1990s, researchers have made vigorous efforts to develop and validate how to use synchrophasor measurements to detect voltage stability problems in real time \cite{Glavic11}. However, these approaches are based on a corridor with a single line, and there are difficulties in directly applying the methods to corridors with multiple transmission lines. The present study addresses this difficulty by proposing a way to combine synchrophasor measurements that approximately reduce a corridor with multiple lines to an equivalent single line, giving a more justifiable and accurate indication of the margin to voltage collapse.

In this paper, we apply the new concept of the voltage across an area that is described in \cite{DobsonPS12volt}. The area voltage is a single complex number describing the voltage between the generators and loads that is consistent with circuit laws. Explicitly, in this paper, the equivalent voltage across a single line is obtained by regarding the multiple transmission lines as an area of the power system, through which the generators supply the loads, and then combining the voltage measurements to obtain the voltage across the area. 

The paper is structured as follows. Section II reviews voltage margin monitoring for a single line, and suggests using measurements at both ends of the line. 
Section III shows how to combine the synchrophasor measurements and reduce the system to a single line. Results for a simple power system example are presented in Section IV, and Section V concludes the paper.

\section{Monitoring voltage margin on a single line}

There is extensive previous work about monitoring the voltage collapse margin in real time on a single line with synchrophasor measurements near the load; see \cite{Glavic11, Begovic99}, and references therein.  These single line methods can apply nicely in the simpler radial situations. 

Assuming that the load is constant power, the voltage collapse occurs as a condition of maximum power transfer to the load at which the load impedance and the Th\'evenin impedance of the line viewed from the load are equal in magnitude. The load impedance can be estimated directly from the synchrophasor measurements, and multiple synchrophasor measurements at closely spaced times are made to estimate the Th\'evenin impedance, using iterative methods (least squares \cite{Begovic99}, recursive least square \cite{Milosevic03}, Tellegen's theorem \cite{Gubina06}), and non-iterative methods  \cite{BaiBegovic13}.
Another approach replaces some of the PMU measurements with Th\'evenin impedances calculated from  line status information from the state estimator \cite{DuongEPECS13}.

The timing of the successive measurements needs to be small enough that the Th\'evenin impedance remains sufficiently constant, and needs to be large enough that changes in the load make the successive measurements change enough. Fulfilling these conditions can sometimes pose problems or require sophisticated corrections, especially during transient changes in the Th\'evenin equivalent caused by generation and transmission events \cite{Genet08}.

To avoid these problems, we follow \cite{LarssonBPT03} in making simultaneous measurements at both the generation and load ends of the transmission line. Fig.~\ref{Figuresingleline} shows the single line case:
 \begin{figure}[H]
\centering
\includegraphics [width=0.5\textwidth]{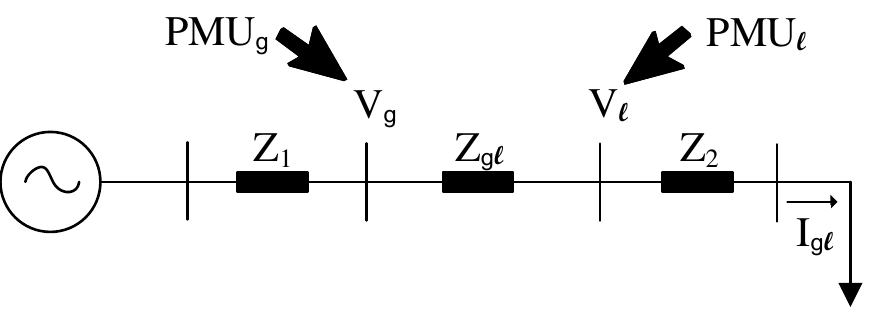}
\caption{
\label{Figuresingleline}%
Single line system }%
\end{figure}
First, it is important to clarify that this approach models the generator as PV, and as  assumed in all the previous methods, the load is modeled as PQ. For this reason the state vector of the system is $x=\{\delta_{g},V_{\ell},\delta_{\ell}\}$, and the  parameters are $\lambda=\{P_{\ell},Q_{\ell}\}$. The parameter $\lambda$ is supposed to vary more slowly than the dynamics, and the dynamics of the parameters are neglected.  The constant power assumption of PQ loads makes the voltage index conservative, underestimating the margin.  

We combine the  synchrophasor complex voltages and currents  to estimate the Th\'evenin impedance according to 
\begin{align}
 Z_{g\ell}&=\frac{V_g-V_{\ell}}{I_{g\ell}}\\
 Z_{\ell}&=\frac{V_\ell}{I_{g\ell}}
 \end{align}
Our intention is to apply this methodology to large transmission corridors with long lines of substantial impedance, where the corridor is defined as a set of transmission lines between the generators and load. Then the impedance $Z_1$ between the corridor and generation, and the impedance $Z_2$ between the corridor and the loads are relatively small, so that the equivalent impedance of the transmission corridor is approximately equal to the Th\'evenin impedance.

\section{Combining measurements }
 
\subsection{Reducing a two input--two output system}
\label{reduce}
 
\begin{figure}[H]
\centering
\includegraphics [width=0.5\textwidth]{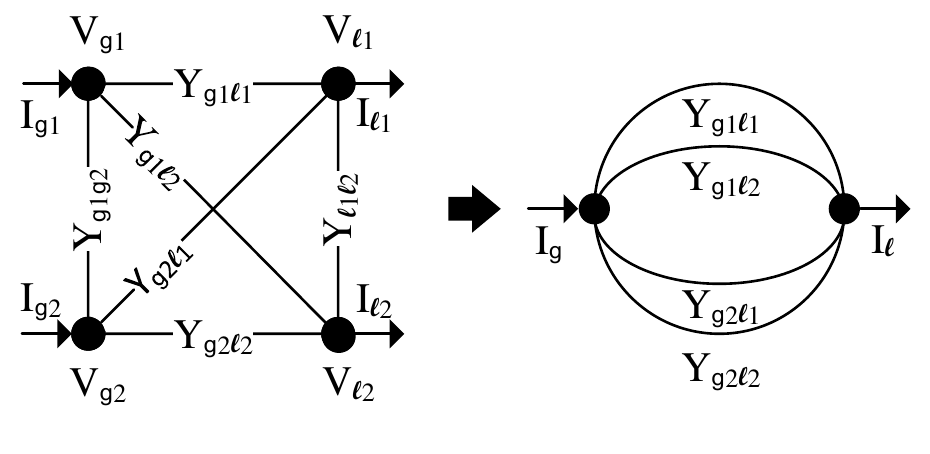}
\caption{
\label{Figurereduce}%
Reduction of power system with 2 inputs and 2 outputs. }%
\end{figure}

As was mentioned before, voltage collapse monitoring with synchrophasors for a single line is very convenient for radial systems. However, in the real world, the power system is usually not radial. In real systems, there are several transmission lines that connect generators to  load centers in a somewhat meshed fashion in order to satisfy their power and reliability requirements. 
Thus, one of the challenges of this paper is to combine several transmission lines into an equivalent single line where the single line method for voltage stability can be applied with confidence. In this section,  we analyze a system with two inputs (generators at bus $g1$ and $g2$) and two outputs (loads
at buses $\ell1$ and $\ell2$), see Fig.~\ref{Figurereduce}, that will be reduced to a system with one input and one output. 
Power networks have many degrees of freedom in their voltages and currents. Here we focus on one degree of freedom in which the generator voltages are equal, the load voltages are equal, and the total current into the transmission corridor is preserved. 
This transmission corridor degree of freedom describes the overall transfer of power from the generation to the load through the transmission corridor. For this degree of freedom of the network, we suppose that the equal generator voltage at buses $g1$ and $g2$ is $V_g$,  the equal load voltage at buses $\ell1$ and $\ell2$ is $V_\ell$, and the total current into the transmission corridor is $I_g$. 

For the corridor degree of freedom, the equal voltages at buses $g1$ and $g2$ imply that buses $g1$ and $g2$ can be contracted or merged into a single bus $g$.  Similarly, equal voltages at buses $\ell1$ and $\ell2$ imply that buses $\ell1$ and $\ell2$  can be contracted into a single bus $\ell$.
Since the transmission lines in the corridor are now in parallel between  bus $g$ and bus $\ell$, the admittance between bus $g$ and bus $\ell$ is 
\begin{align}
Y_{g\ell}=Y_{g1\ell1}+Y_{g1\ell2}+Y_{g2\ell1}+Y_{g2\ell2},
\label{Ygl}
\end{align}
and it follows (see Fig.~\ref{Figurereduce}) that 
\begin{align}
I_g=Y_{g\ell}(V_g-V_\ell).
\label{Ig}
\end{align}
Now the total current entering the transmission corridor is 
\begin{equation}
\begin{aligned}
I_{g1}+I_{g2}&=(V_{g1}-V_{\ell1})Y_{g1\ell1}+(V_{g1}-V_{\ell2})Y_{g1\ell2}\\
&\quad+(V_{g2}-V_{\ell1})Y_{g2\ell1}+(V_{g2}-V_{\ell2})Y_{g2\ell2}.
\label{Ig1plusIg2}
\end{aligned}
\end{equation}
We want the current (\ref{Ig}) entering the corridor in the corridor degree of freedom to be equal to the total current (\ref{Ig1plusIg2})  entering the corridor in the entire network:
\begin{align}
I_g=I_{g1}+I_{g2},
\label{Igequal}
\end{align}
which can also be expressed as
\begin{equation}
\begin{aligned}
Y_{g\ell}(V_g-&V_\ell)=
(V_{g1}-V_{\ell1})Y_{g1\ell1}+(V_{g1}-V_{\ell2})Y_{g1\ell2}\\
&+(V_{g2}-V_{\ell1})Y_{g2\ell1}+(V_{g2}-V_{\ell2})Y_{g2\ell2}.
\label{Igequal2}
\end{aligned}
\end{equation}
Now we use (\ref{Igequal2}) to determine the values of $V_g$ and $V_\ell$. As we are only interested in one degree of freedom, the generation voltage in the complete system should be equal to the generation voltage in the reduced system; and similarly for the load buses. Namely, separating the terms of the equation (\ref{Igequal2}) according to generation and load, we can solve (\ref{Igequal2}) with
\begin{align}
\label{Vgen}
V_{g}&= w_{g1}V_{g1}+w_{g2}V_{g2}\\
V_{\ell}&=w_{\ell1}V_{\ell1}+w_{\ell2}V_{\ell2},
\label{Vload}
\end{align}
where the weights are defined as 
\begin{align}
\begin{aligned}
w_{g1}&= \frac{Y_{g1\ell1}+Y_{g1\ell2}}{Y_{g\ell}},\qquad
w_{g2}= \frac{Y_{g2\ell1}+Y_{g2\ell2}}{Y_{g\ell}},
\label{w1}\\
w_{\ell1}&=\frac{Y_{g2\ell1}+Y_{g2\ell1}}{Y_{g\ell}},\qquad
w_{\ell2}=\frac{Y_{g2\ell2}+Y_{g2\ell2}}{Y_{g\ell}}.
\end{aligned}
\end{align}
Then the complex voltage difference across the combined transmission lines is 
\begin{align}
V_{g\ell}= V_{g}-V_{\ell}.
\label{Vgl}
\end{align}
If we regard the transmission lines as an area of the power system, then $V_{g\ell}$ is the complex voltage across the area described in \cite{DobsonPS12volt}
(compare (\ref{Vgen}-\ref{Vgl}) with equation (4) of \cite{DobsonPS12volt}). This approach only depends on the admittances of the lines in the area, and the voltages at the boundaries of the area are combined according to the weights of these admittances. 

Equations (\ref{Vgen}) and (\ref{Vload})  combine the voltages of the generation buses and combine the voltages of the  load buses,  resulting in a reduced system which is a compound of  the combination of the voltages, the admittances of the transmission lines that connect the load with the generation, and the sum of the currents in the generation buses  and the load buses.

The reduction above applies to the transmission corridor degree of freedom that transfers power from the generators to the loads. If the power network has equal voltages at the generators, and equal voltages at the loads, then only the transmission corridor degree of freedom is present in the network currents and voltages, and the reduction applies perfectly to the power system.
In this case, the contraction of the generator buses into a single bus and the contraction of the load buses into a single bus do not affect the behavior of the network. In particular, the maximum power transfers of the original and reduced networks must be identical. 

When the generator voltages differ or the load voltages differ, there are additional degrees of freedom of the network present in the network currents and voltages, and the maximum power transfers of the original and reduced networks can differ. We investigate the impact of this difference on our methods numerically in section \ref{imperfect}.

In practical high stress cases of interest, we expect that lines with larger susceptances carry larger currents and powers, so that the voltage across the lines can be similar, and the equal voltage conditions at the generators and at the loads can be approximately satisfied.

 \subsection{Synchrophasor monitoring of two input-two output system}
 \begin{figure}[H]
\centering
\includegraphics [width=0.45\textwidth]{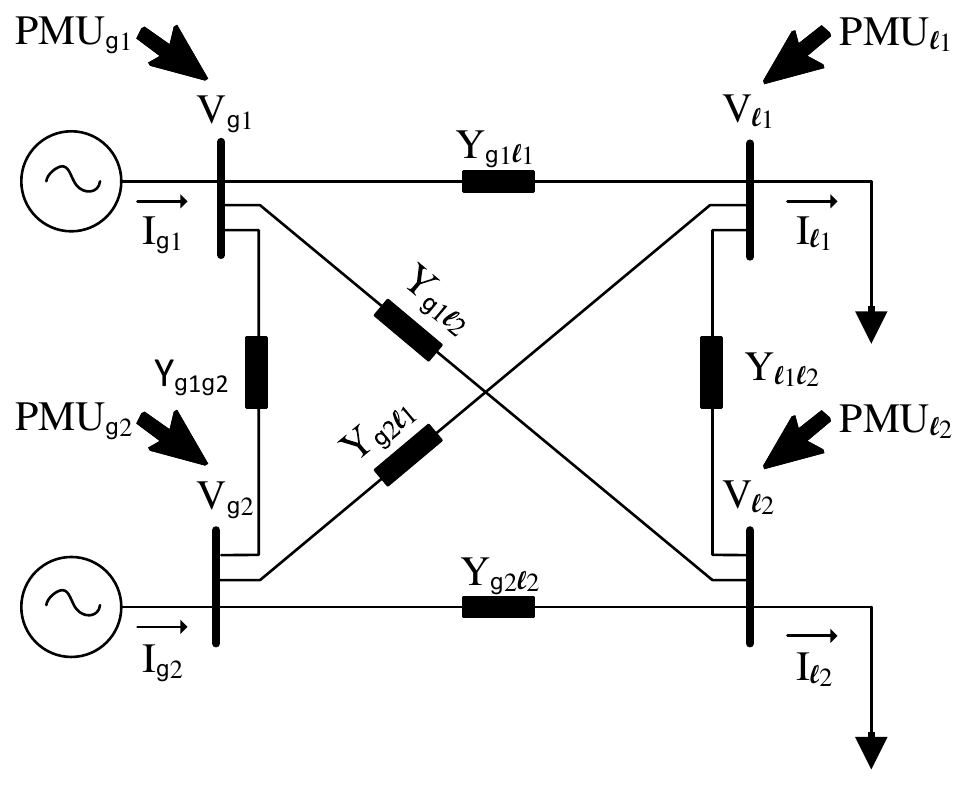}
\caption{
\label{Figure2corridor}%
Power system with 2 inputs and 2 outputs and 4 lines in corridor.}%
\end{figure}

\begin{figure}[H]
\centering
\includegraphics [width=0.45\textwidth]{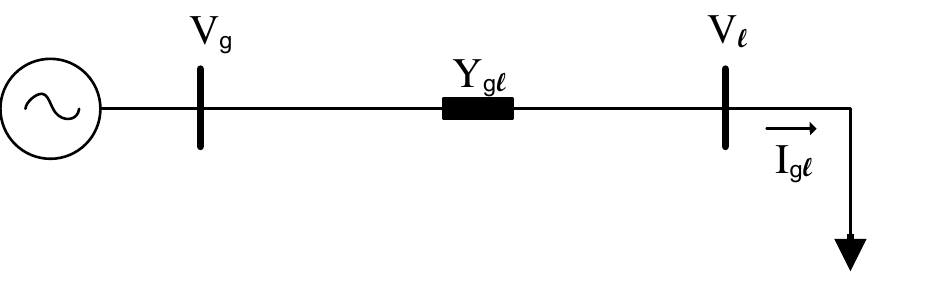}
\caption{
\label{Figurereducesingleline}%
Reduced power system.}%
\end{figure}

At this point we have a way to reduce multiple transmission lines to a single line, but to apply this in real time it is necessary to use the synchrophasor measurements. 
This requires 3 steps:

\begin{enumerate}
\item Measure the complex voltage and current at both ends of all the transmission lines in the corridor, see  Fig.~\ref{Figure2corridor}.
\item Use the synchrophasor measurements to find the admittance and weight of each transmission line.
\item Combine the line admittances into the equivalent corridor admittance; combine the  measurements of complex current and voltage into the 
equivalent voltage across and current through the corridor, see Fig.~\ref{Figurereducesingleline}.
\end{enumerate}
The case that is analyzed in this paper has two input or generators, modeled as PV; and the outputs are two loads, modeled as PQ. Now we follow the previous 3 steps.

The admittances of each transmission line, calculated using the synchrophasor measurements of its complex current and the complex voltages across the line, are
\begin{align*}
Y_{g1\ell1}=\frac{I_{g1\ell1}}{V_{g1}-V_{\ell1}},\qquad\quad
Y_{g1\ell2}=\frac{I_{g1\ell2}}{V_{g1}-V_{\ell2}},\\
Y_{g2\ell1}=\frac{I_{g2\ell1}}{V_{g2}-V_{\ell1}},\qquad\quad
Y_{g2\ell2}=\frac{I_{g2\ell2}}{V_{g2}-V_{\ell2}}.
\end{align*}
The synchrophasor measurements of voltage are combined using the weights (\ref{w1}). In this way, the reduced generator bus voltage, load bus voltage, voltage across the reduced system, and current of the reduced system are 
\begin{align}
&V_{g}= w_{g1}V_{g1}+w_{g2}V_{g2}\\
&V_{\ell}=w_{\ell1}V_{\ell1}+w_{\ell2}V_{\ell2}\\
&V_{g\ell}=  w_{g1}V_{g1}+w_{g2}V_{g2}-w_{\ell1}V_{\ell1}-w_{\ell2}V_{\ell2}\\
&I_{g\ell}=I_{\ell1}+I_{\ell2}
\end{align}.

 \subsection{Voltage stability margin}

One stability index using the Th\'evenin equivalent that can be found in the literature is based on tracking the absolute value of the load impedance \cite{Begovic99}. The maximum transfer of power occurs when the absolute value of the impedance of the load is equal to the absolute value of the Th\'evenin impedance:
\begin{align}
\mid{Z_{load}\mid= \mid{Z_{TH}}}\mid.
\end{align}
Another common index of voltage stability,
\begin{align}
\displaystyle\frac{V_{\ell}}{V_{TH}-V_{\ell}},
\end{align}
 is based on the fact that the maximum transfer of power occurs when the voltage of the load bus is equal to the voltage drop in the Th\'evenin impedance \cite{Milosevic03}.

In this paper, we propose monitoring the proximity to the maximum power transfer and voltage collapse with an index based on the apparent power. The formulation of the index in terms of the apparent power should help to make the index clear to the operators. 
\begin{align}
\mbox{Index}&=\frac{ \mid{S_{g\ell}\mid100}}{{ \mid{S_{\ell}}} \mid},
\label{VSIS}\\
%
%
\mbox{where}\quad&S_{g\ell}=V_{g\ell}{I_{g\ell}}^*\\
&S_{\ell}= V_{\ell}{I_{g\ell}}^*.
\end{align}
Index (\ref{VSIS}) indicates the percentage of the maximum load apparent power that can be achieved under the measured condition of the corridor. 
An alarm can be triggered when a sufficient percentage of the maximum load apparent power is exceeded.  For example, the alarm could be triggered when the index exceeds 80\%.

 \section{Illustration and test of the reduction }
 
 \subsection{Perfect reduction case}
\begin{figure}[t]
\centering
\includegraphics [width=0.5\textwidth]{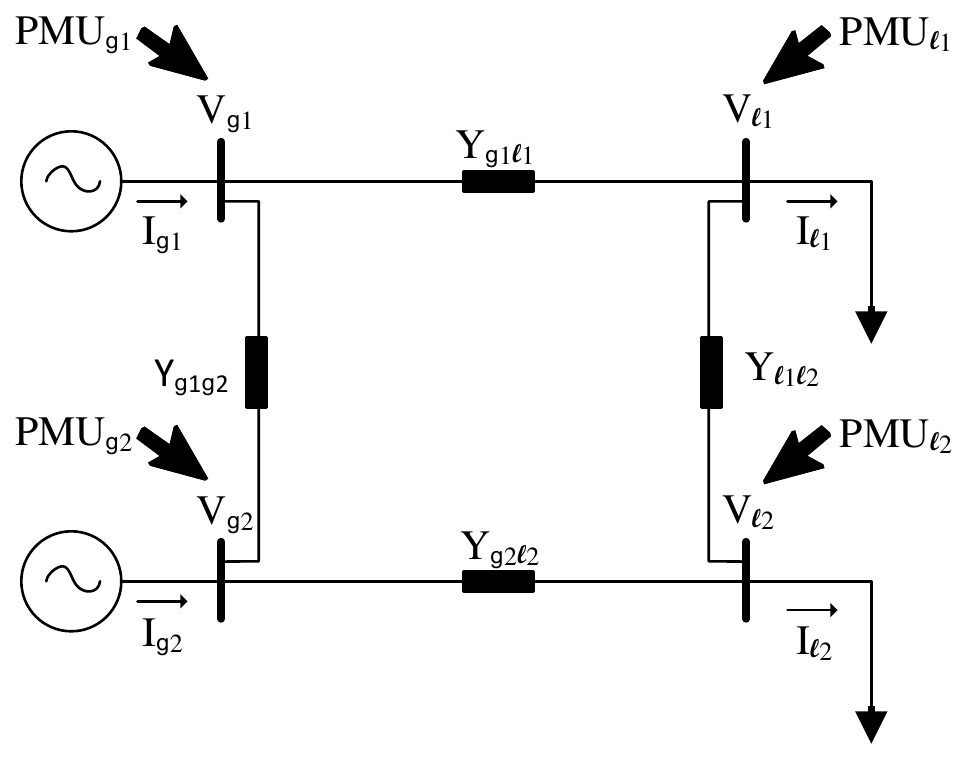}
\caption{
\label{Figuretest}%
Power system with 2 inputs and 2 outputs and 2 lines in corridor.}%
\end{figure}
As explained in section \ref{reduce}, the reduction, and hence the performance of the index, is perfect when the generation voltages are equal and the load voltages are equal. 
This subsection illustrates this case numerically. We choose equal generation and equal load voltages, shown in Table~\ref{example}, for the four bus system  of  Fig.~\ref{Figuretest}. The values for this power system and its reduction have been adjusted to be at the maximum power 
voltage collapse condition, and the results in Table~\ref{example} show that the maximum power transfer is the same for both the complete and reduced systems.
\begin{table}[H]
\centering
\caption{Example of perfect reduction}
\label{example}
\vspace{-2pt}
\begin{tabular}{llllllll}
 \hline\\[-6pt]
 \multicolumn{2}{c}{Maximum power transfer of the complete system\rule[-5pt]{0pt}{0pt}}\\
$V_{g1}=1+j0$&$V_{g2}=1+j0$ \\[1pt]
$V_{\ell1}=0.5-j0.2$&$V_{\ell2}=0.5-j0.2$&\\[1pt]
$I_{g1\ell1}=6.1 -j8.7$ &$I_{g2\ell2}=18.4-j26.1$&\\[1pt]
$I_{g1g2}=0$ &$I_{\ell1\ell2}=0$&\\[1pt]
$Y_{g1\ell1}=3.8 - j19.1$&$Y_{g2\ell2}=11.4 - j57.2$\\[1pt]
$Y_{g1g2}=5.2 - j25.8$&$Y_{\ell1\ell2}=8.2 - j34.8$\\[1pt]
$S_{max}=20 - j 12$ \\[2pt]
 \hline\\[-6pt]
 \multicolumn{2}{c}{ Maximum power transfer of the reduced system \rule[-5pt]{0pt}{0pt}}\\
$w_{1}=0.25$&$w_{2}=0.75$&\\[1pt]
$V_{g}=1+j0$&$V_{\ell}=0.5-j0.2$ \\[1pt]
$V_{g\ell}=0.5 + j 0.2$&$I_{g\ell}=24.5 - j34.8$\\[1pt]
$Y_{g\ell}=15.3 - j76.3$&$Y_{\ell}=66.73 - j40.04$\\[1pt]
$\mid{Y_{g\ell}}\mid=77.82$&$\mid{Y_{\ell}}\mid=77.82$\\[1pt]
$S_{g\ell}=4.6+ j 22.8$&$S_{\ell}=20+ j 12$\\[1pt]
$\mid{S_{g\ell}} \mid=23.3$&$\mid{S_{\ell}} \mid=23.3$\\[1pt]
Index$=100\%$\\[1pt]
\hline\\[-6pt]
\end{tabular}
\end{table}

 \subsection{General reduction case and numerical assessment of errors }
 \label{imperfect}
 
When the generation bus voltages are not equal and/or  the load bus voltages are not equal, the maximum transfers of power in the complete and reduced system are not  exactly the same,
and the index will have an error. This subsection measures this error numerically to help judge its significance.

In the four bus system of  Fig.~\ref{Figuretest}, the generation voltages were assumed to be equal, the load power factor was held constant,  and we varied the load voltages by 
transferring power between the loads.

For each set of load voltages, the error in the maximum real load power was assessed using the following steps:
\begin{enumerate}
\item Evaluate the maximum transfer of power for the complete system by increasing  load powers proportionally.
\item Use the total load power of the complete system as the initial load power in the reduced system.
\item Evaluate the maximum transfer of power for the reduced system by varying the load power.
\item Estimate the error by comparing the maximum total load power of the complete system with the maximum load power of the reduced system.
\end{enumerate}
The results in Table~\ref{error} show that the error in the maximum transfer of real power exceeds 10\%  when the difference of the load voltage magnitude exceed 20\%, and the difference between the voltage angles are more than $13^{\circ}$. 
In practical power systems, the load voltage magnitude would vary less than 10\% and the difference of the voltage angles would vary less than $10^{\circ}$.

This example suggests that the voltage margin index inaccuracy due to the reduction should be less than 10\% in the more extreme practical cases, and the inaccuracy due to the reduction would often be much smaller for less extreme cases.
\begin{table}[H]
\centering
\caption{Maximum power transfer for a system with two corridors}
\label{error}
\vspace{-4pt}
\begin{tabular}{ccccc|cc}
\multicolumn{5}{c|}{Complete system}& \multicolumn{2}{c}{Reduced system}\\[4pt]
$P^{\mathsf{max}}_{\ell1}$&$P^{\mathsf{max}}_{\ell2}$&$P^{\mathsf{max}}_{\ell}$&$V_{\ell1}\!-\!V _{\ell2}$&$\delta_{\ell1}\!-\!\delta_{\ell2}$&$P^{\mathsf{max}}_{\ell}$&error\\[4pt]
 \hline\\[-8pt]
9.7&0.0&9.7&-0.3&-19&10.6&-10\\[1pt]
9.5&1.1&10.6&-0.29&-18&12.5&-18\\[1pt]
9.2&2.3&11.5&-0.28&-18&14.3&-24\\[1pt]
9.0&3.8&12.8&-0.26&-17&15.8&-23\\[1pt]
8.6&5.7&14.3&-0.24&-16&17.2&-20\\[1pt]
8.0&8.0&16&-0.21&-14&18.4&-14\\[1pt]
7.7&9.4&17.1&-0.19&-13&19.0&-11\\[1pt]
7.2&10.9&18.1&-0.16&-11&19.4&-7\\[1pt]
5.9&13.8&19.7&-0.06&-5&19.9&-1\\[3pt]
\bf 5.0&\bf14.9&\bf19.9&\bf0&\bf0&\bf19.9&\bf0\\[3pt]
4&15.8&19.8&-0.05&4&19.9&-1\\[1pt]
1.9&17.0&18.9&0.13&10&19.3&-2\\[1pt]
0.2&17.8&18&0.17&13&18.3&-2\\[1pt]
\hline\\[-8pt]
\multicolumn{7}{c}{ Angles in degree, error in percent, and all other quantities in per unit.}\\[1pt]
\multicolumn{7}{c}{ Perfect reduction case has zero error and is shown in bold face.}
\end{tabular}
\end{table}

\section{Conclusion}
\label{conclusion}

It is advantageous to supplement voltage collapse margin monitoring that depends on the state estimator with 
real time monitoring with synchrophasors. This can give a fast warning to operators and can function even if the state estimator does not converge.
In this paper, we propose to monitor voltage instability by combining synchrophasor measurements at both ends of 
several transmission lines that join generators to loads. The case of two generators and two loads is considered.

The synchrophasor measurements are combined into an equivalent voltage between the generation and load 
by regarding the transmission lines as an area of the power system and applying the concept of the voltage across the area.  The voltage across an area describes a degree of freedom of the network that supplies power from the generators to the loads. We informally derive the area voltage in this paper 
and refer to \cite{DobsonPS12volt} for a more thorough derivation.

The combining of the synchrophasor measurements using 
the voltage area concept  has the  effect of  reducing  multiple transmission lines to a single line equivalent to which on-line voltage stability monitoring with synchrophasor measurements can be applied. Our monitoring of the single line equivalent is similar to previous work except that we use synchrophasor voltages at both ends of the line and express the resulting voltage margin index in terms of complex power.

The reduction to a single line equivalent is perfect in the special case in which the loads have equal voltages and the generators have equal voltages.
That is, in this case, the on-line monitoring of the equivalent single line exactly gives the margin to voltage collapse for the original system.
This is a special case, but we studied the general case with a numerical example and found that  the errors were reasonably small. 
 
Our results suggest that we have found a promising and systematic approach for combining together synchrophasor measurements for multiple 
transmission lines so that single line monitoring methods can be applied. For future work we will generalize and further test the approach, analyze why the equal voltage approximation works so well, and consider implementation issues in a  control center.
One important topic for future work is testing or extending the method to handle generator reactive power limits, and some previous approaches are in \cite{Glavic11,Milosevic03,DuongEPECS13}.

\vspace{10pt}

\section{Acknowledgements}
\label{ack}
We gratefully acknowledge support in part from the Electric Power Research Center at Iowa State University, 
DOE project ``The Future Grid to Enable Sustainable Energy Systems," an initiative of PSERC, 
Arend J. and Velma V. Sandbulte professorship funds, and
NSF grant CPS-1135825.

\end{document}